\documentstyle[aps,preprint]{revtex}
\draft
\begin{document}
\title {Use of synchronization and adaptive control in parameter
estimation from a time series} 
\author{Anil Maybhate\cite{em1} $^{1,2}$ and R. E. Amritkar\cite{em2}$^1$}
\address{$^1$Physical Research Laboratory, Navrangpura, Ahmedabad 380009 India} 
\address{$^2$Department of Physics, University of Pune, Pune 411007 India}
\maketitle

\begin{abstract} A technique is introduced for estimating unknown
parameters when time series of only one variable from a multivariate
nonlinear dynamical system is given. The technique employs a combination
of two different control methods, a linear feedback for synchronizing
system variables and an adaptive control, and is based on dynamic
minimization of synchronization error. The technique is shown to work
even when the unknown parameters appear in the evolution equations of the
variables other than the one for which the time series is given. The
technique not only establishes that explicit detailed information about
all system variables and parameters is contained in a scalar time series,
but also gives a way to extract it out under suitable conditions.
Illustrations are presented for Lorenz and R\"ossler systems and a
nonlinear dynamical system in plasma physics. Also it is found that the
technique is reasonably stable against noise in the given time series and
the estimated value of a parameter fluctuates around the correct value,
with the error of estimation growing linearly with the noise strength, for
small noise. \end{abstract}

\section{introduction}

One of the objectives of time series analysis is to study the detailed
structure of the equations of the underlying dynamical system which govern
its temporal evolution. This includes the number of independent variables,
the form of the flow functions, the nonlinearities in them and parameters
of the system~\cite{AB}. This paper concentrates on the last aspect, i.e.,
estimating the parameters of a nonlinear system from a single time series
when partial information about the system dynamics is
available~\cite{PJ,Par,PKSP}.

Assuming that the number of independent variables and the structure of
underlying dynamical evolution equations for a nonlinear system is known,
we address the problem of determining the values of the parameters. In
particular, given a time series for a single variable (a scalar time
series), we suggest a simple method which enables us to determine values
of the unknown parameters dynamically. The unknown parameters may or may
not appear in the evolution equation of the variable for which the time
series is given. For this purpose, we employ a combination of two
techniques, namely synchronization and adaptive control.

Owing mainly to the extreme sensitivity to initial conditions, engineering
and controlling a nonlinear chaotic system requires a careful analysis.
Feedback based synchronization techniques are investigated in this context
to force a chaotic system, to go to a desired periodic or chaotic orbit.  
Such control mechanisms were suggested by Pecora and Carroll
~\cite{PC1,PC2} and many others~\cite{Sin1,Sin2,CD,Pyr,PC3,JA1,JA2} with
an aim to synchronize two chaotic orbits and to stabilize unstable
periodic orbits or fixed points. In such mechanisms, some of the
independent variables are used as drive variables and the remaining
variables are found to synchronize with the desired trajectory under
suitable conditions. There have been many other important attempts in
controlling chaotic systems using
synchronization~\cite{MH,LG,LP,BG,Ani,Rul,Koc}.

The other method that we use is that of adaptive control which is used to
bring back a system, deviated from a stable fixed point due to changes in
parameters and variables, to its original state. This mechanism was
suggested by Huberman and Lumer~\cite{HL}. It was generalized for an
unstable periodic orbit and a chaotic orbit by John and
Amritkar~\cite{JA1} where it was shown that it is possible to synchronize
with an unstable periodic orbit or a chaotic orbit starting from a random
initial condition and different value of the parameter.

In this paper, we show that a simple combination of synchronization and
adaptive control methods similar to that described by John and
Amritkar~\cite{JA1,JA2} can be used for extracting information contained
in a scalar time series.

We approach the problem by considering a dynamical system, in which the
number of independent variables and the structure of evolution equations
are assumed to be known. A linear feedback function is added to the
variable corresponding to that for which the time series is given. This
acts as a {\it drive} variable. The feedback serves the purpose of
synchronization of all the system variables. The feedback function in our
case is proportional to the difference between the new and the old values
of the drive variable.

The system variables respond to this feedback by synchronizing with the
corresponding values in the original system. In the context of application
of synchronization techniques to telecommunications, the new system to be
reconstructed is often referred to as the {\it receiver} whereas the old
system, from which the time series is made available is termed the {\it
transmitter}. We will borrow the terminology, although the meanings of
terms in the two cases are not exactly identical.

The synchronization as described above becomes exact when the receiver
parameters are set equal to those of the transmitter and takes place
whenever the {\it Conditional Lyapunov Exponents} (CLE's) as defined in
the next section are all negative. Now assume that precise values of only
a few of the transmitter parameters are known to the receiver system. We
show that, in such a case it is possible to write simple evolution
equations for the unknown parameters (initially set to arbitrary values),
which when coupled with the system equations, yield precise values of all
the state variables and the unknown parameters asymptotically to any
desired accuracy.

Our method comprises of raising the unknown parameters to the status of
variables of a higher dimensional dynamical system which evolve according
to a simple set of evolution equations. The receiver forms a subsystem of
this higher dimensional system which in addition contains the evolution
equation for the unknown parameters. The input to this higher dimensional
system is a scalar time series obtained from the transmitter system. Thus
our method uses a dynamical algorithm to estimate the parameters which are
obtained asymptotically. We note that the method of estimating parameters
using synchronization and minimization as proposed in Ref.~\cite{PJ}, is
essentially a static method. The problem of estimating model parameters
was also handled in Ref.~\cite{Par}, in which starting with an ansatz, the
optimal equations for parameter evolution are obtained. Our method gives a
simpler and a systematic derivation of the parameter control loop and in
many cases, a better convergence rate.

It is well known that a great deal of information about a chaotic system
is contained in the time series of its variables. Techniques like
embedding the time series in a space with chosen dimensionality are
available for studying the universality class and other {\it global}
features of the system. Our results suggest that a scalar time series, in
addition to the information about the universality class also contains
information about the exact values of the parameters of the underlying
dynamical system, including the ones which appear in the evolution of
other variables.

The method and the required notation is developed in section II. Section
III consists of illustrations for Lorenz and R\"ossler systems and a set
of equations in plasma physics. The effect of noise in the transmitter
system is studied in Section IV. Finally we conclude in section V with a
brief summary of results along with a few remarks.

\section{formalism} 
\subsection{Description of the method}
 
In this section, we will describe our method of parameter estimation, for
a general system with $n$ variables and $m$ parameters. We will first
consider the case when only one parameter is unknown to the receiver.

Consider an autonomous, nonlinear dynamical system with evolution
equations,
\begin{eqnarray} 
{\bf\dot x} = {\bf f}\left({\bf x},\{\mu_j\}\right),
\label{transmitter} 
\end{eqnarray}
where ${\bf x} = (x_1,x_2,\ldots,x_n)$ is an $n$-dimensional state vector
whose evolution is described by the function ${\bf f}=(f_1,\ldots,f_n)$.  
The dot represents time differentiation and $\{\mu_j\}, j=1,2,\ldots,m,$
are the parameters of the system.

Now suppose a time series for one variable, which without loss of
generality can be taken as $x_1$, is given as an output of the above
system and in addition suppose the functional form of $\bf f$, and the
values of all the parameters $\mu_j, j=1,\ldots,l-1,l+1,\ldots,m,$ are
known while the time evolution of the remaining variables and value of
${\mu}_l$, the $l^{th}$ parameter are not known, then formally the problem
at hand consists of writing a set of evolution equations which will yield
the information about the unknown parameter and also other variables. With
the unknown parameter ${\mu}_l$ written explicitly for convenience we
rewrite Eq.~(\ref{transmitter}) as,
\begin{eqnarray} 
{\bf\dot x} = {\bf f}\left({\bf x},\{\mu_j|j\ne l\},\mu_l\right).
\label{transmit}
\end{eqnarray}

Now we introduce a new system of variables ${\bf x'}=(x'_1,x'_2,\ldots,x'_n)$ 
whose evolution equations have identical form to that of $\bf x$. We fix
$x'_1$ as the drive variable and a feedback is introduced in the evolution
of $x'_1$. The parameters are also the same except the one corresponding
to the unknown parameter which will be set to an arbitrary initial value
denoted by $\mu'_l$. Thus the receiver system will have the structure,
\begin{eqnarray} 
\dot x'_1 &=& g\left( {\bf x'} , \{\mu_j|j\ne l\}, \mu'_l \right)\nonumber
\\
&=& f_1 \left( {\bf x'} , \{\mu_j|j\ne l\}, \mu'_l \right) 
- w_f \left( x'_1, x_1(t) \right),\label{drive}
\\
\dot x'_i &=& f_i \left( {\bf x'}, \{\mu_j|j\ne l\}, \mu'_l \right),
\;\;\;i=2,\ldots,n,
\label{receive}
\end{eqnarray}
where $w_f(x'_1,x_1(t))$ is a feedback function which depends upon the
drive variable $x'_1$ and the variable $x_1$. The feedback function can be
most simply chosen to be proportional to the difference $(x'_1 - x_1)$ and
the evolution for the drive variable $x'_1$ can be written as,
\begin{eqnarray}
\dot x'_1 = f_1\Big( {\bf x'} , \{\mu_j|j \ne l\} , \mu'_l \Big) 
- \epsilon \left(x'_1 - x_1(t)\right), 
\label{feedback}
\end{eqnarray}
where $\epsilon$ is called the feedback constant. More general forms
of the feedback function are also possible and give similar results.

The receiver system is formed by Eqs.~(\ref{receive}) and
(\ref{feedback}). If the parameter $\mu'_l$ in these equations is set
precisely equal to $\mu_l$ then the two sets of variables, ${\bf x}$ and
${\bf x'}$, after a transient time, evolve in tandem and show exact
synchronization under suitable conditions, but because the value of
$\mu_l$ is {\it unknown} to the receiver system, this does not happen.

The solution is to set the parameter $\mu'_l$ to an arbitrary initial
value, while all others are set to the known values $\mu_j$, and {\it
adapt} it through a suitable evolution equation. The resulting
$(n+1)$-dimensional system then evolves all the receiver variables to
correct values of the corresponding transmitter variables and
simultaneously settles the value of $\mu'_l$ to that of $\mu_l$ provided
all the CLE's as defined in the next subsection, are negative.

The equation for evolution of the $\mu'_l$ is chosen similar to those used
in adaptive control mechanisms~\cite{JA1,JA2}, and quite generally can
have the form,
\begin{eqnarray} 
\dot \mu'_l = h\left( \left( x'_1 - x_1(t) \right),
{\partial g\over\partial\mu'_l}\right).
\label{genadapt}
\end{eqnarray}
The form of the function $h$ that we have chosen is
\begin{eqnarray} 
\dot \mu'_l = -\delta \big(x'_1 - x_1(t)\big)
{\partial g\over\partial\mu'_l}, 
\label{adaptation}
\end{eqnarray}
where $\delta$ is another parameter in the combined $(n+1)$~-dimensional
system formed by Eqs.~(\ref{receive}), (\ref{feedback}) and
(\ref{adaptation}).
We call it the {\it stiffness constant}. The values
of $\epsilon$ and $\delta$ together control the convergence rates involved
in synchronization and adaptive evolution.
Towards the end of this subsection, we will show that the above form
of function $h$ (Eq.~(\ref{adaptation}))
is obtained as a result of dynamic minimization of the synchronization
error.

The last factor in the Eq.~(\ref{adaptation}), $({\partial
g/\partial\mu'_l})$, needs some elaboration. In general the parameter
$\mu'_l$ may or may not explicitly appear in the evolution function
$g({\bf x'},\{\mu_j|j\ne l\},\mu_l')$ in Eq.~(\ref{drive}). This stresses
a need for identification of two separate cases.

If the function $g$ explicitly depends on $\mu'_l$, then the calculation
of $({\partial g/\partial\mu'_l})$ is straightforward.
 
In case $\mu'_l$ does not appear in the function $g$ explicitly, it still
indirectly affects the evolution of $x'_1$. The information about the
value of $\mu_l$ is contained in the given time series $x_1(t)$. Function
$({\partial g/\partial\mu'_l})$ ``taps'' this dependence. The calculation
of $({\partial g/\partial\mu'_l})$ in this case, when function $g$ does
not explicitly depend on $\mu'_l$ needs to be done carefully. This is done
as follows :

Consider the system formed by (\ref{receive}) and (\ref{feedback}) in
which, a change in the variable $x'_1$ in one time step
due to a change in the parameter $\mu^{'}_l$ can be estimated as follows.
\begin{eqnarray} 
\Delta x'_1 &\approx&  \Delta g\;dt \nonumber  
\\ 
&\approx&  {\partial g \over \partial x'_s}\Delta x'_s \;dt \nonumber 
\\ 
&\approx&  {\partial g \over \partial x'_s}\Delta f_s  \;(dt)^2  \nonumber
\\ 
&\approx&  {\partial g \over \partial x'_s}{\partial f_s \over
            \partial\mu'_l} \Delta\mu'_l \;(dt)^2,  \nonumber
\end{eqnarray} 
where $x'_s$ is the $s^{th}$ variable of the receiver, such that its 
evolution contains the parameter $\mu'_l$ explicitly. Thus the last
of the above equations gives us, 
\begin{eqnarray}
{\partial g \over\partial\mu'_l} \approx 
{\partial g\over\partial x'_s}{\partial f_s \over\partial\mu'_l}. 
\label{signcaln1}
\end{eqnarray}

A further complication arises if the variable $x'_s$ itself does not
appear in the function $g$ explicitly. In such a case further dependences
appearing in more time steps may be considered. Note that here, $x'_s$ may
appear in more than one flow functions and a summation over all such
functions becomes necessary. In this case we can write,
\begin{eqnarray} 
\Delta x'_1 &\approx&  \Delta g\;dt \nonumber  
\\ 
&\approx&  {\partial g \over \partial x'_s}\Delta x'_s \;dt \nonumber 
\\ 
&\approx&  {\partial g \over \partial x'_s}\Delta f_s  \;(dt)^2  \nonumber
\\ 
&\approx& \left\{ \sum_k {\partial g \over \partial x'_k}{\partial x'_k
\over\partial x'_s} \right\} {\partial f_s \over \partial\mu'_l} \Delta
\mu'_l\;(dt)^2. \nonumber
\\
&\approx&  \left\{ \sum_k {\partial g \over \partial x'_k}{\partial f_k
\over\partial x'_s} \right\} {\partial f_s \over \partial\mu'_l} \Delta
\mu'_l\;(dt)^3. \nonumber
\end{eqnarray} 
Thus the last factor in Eq.~(\ref{adaptation}) takes the form,
\begin{eqnarray}
{\partial g \over \partial\mu'_l} \approx 
\left\{ \sum_k  {\partial g \over \partial x'_k}{\partial
f_k \over \partial x'_s}\right\} {\partial f_s \over \partial \mu'_l}. 
\label{signcaln2}
\end{eqnarray}
One such case appears in the example of Lorenz system, which will be
discussed in the next section. 

Now, when more than one parameters of the transmitter are to be estimated,
one may use a set of equations similar in form to that of
Eq.~(\ref{adaptation}). We will use such a set when we discuss Lorenz
system where it will be assumed that two or three parameters of the Lorenz
system are unknown to the receiver system. We note that a parameter
estimation algorithm as described in Ref.~\cite{Par} can also be used in
the estimation of more than one unknown parameters. It uses
autosynchronization method based on an Active Passive Decomposition (APD)
of a dynamical system~\cite{PKSP} and starts from an ansatz for the
parameter control. In contrast, our method is a dynamical minimization for
the synchronization error. This can be seen as follows :

Let us define the dynamical synchronization error $ e(\mu'_l,t)$ as,
\begin{equation}
e(\mu'_l,t) = (x'_1-x_1)^2,
\label{errdef}
\end{equation}
where $\mu'_l$ is the receiver parameter corresponding to the unknown
parameter and $x'_1$ is the drive variable.

We note that if $\mu'_l$ takes precisely the value of $\mu_l$, then
the transmitter and receiver synchronize, which makes the error as defined
by Eq.~(\ref{errdef}) minimum, i.e. zero. To go to this minimum, we want
to evolve $\mu'_l$ such that it will go to a value making $e(\mu'_l,t)$
minimum. With an analogy to an equation in mechanics, where an overdamped
particle goes to a mimimum of a potential, we write the following,
\begin{equation}
\dot\mu'_l \propto -{\partial e(\mu'_l,t)\over \partial \mu'_l},
\end{equation}
which leads to, 
\begin{eqnarray}
\dot\mu'_l \propto - (x'_1-x_1)  {\partial{x'_1}\over{\partial \mu'_l}},
\label{minimisn0}
\end{eqnarray}
Further, to the lowest order in $dt$, $\Delta x^{'}_1 =
{\partial{\dot{x}^{'}_1}
\over \partial \mu'_l} \Delta \mu^{'}_l \, dt $.
Hence Eq.~(\ref{minimisn0}) can be written as,
\begin{eqnarray}
\dot\mu'_l = - \delta (x'_1-x_1)  {\partial{\dot x'_1}\over{\partial
\mu'_l}}\;,
\label{minimisn}
\end{eqnarray}
where $\delta$ is a proportionality constant. This equation is same as
Eq.~(\ref{adaptation}).

In the next subsection we will define the {\it conditional Lyapunov
exponents} (CLE's) for the newly reconstructed receiver system and state
the condition for the combination of synchronization and adaptive control
to work convergently such that parameter estimation is possible.

\subsection{Condition for convergence}

Consider the transmitter equations (Eq.~(\ref{transmit})) and the
receiver equations (Eqs.~(\ref{receive}), (\ref{feedback}) and
(\ref{adaptation})). Convergence between two trajectories of these systems
means that the receiver variables evolve such that the differences
$(x'_k-x_k), (k=1,\ldots,n)$ and $(\mu'_l-\mu_l)$ all evolve to zero. In
the $(n+1)$-dimensional space formed by these differences, origin acts as
a fixed point and the condition for the algorithm to work is the same as
the stability condition for this fixed point.

If the above differences are considered to form an $(n+1)$-dimensional
vector ${\bf z} =
(z_1,\ldots,z_{n+1})=(x'_1-x_1,\ldots,x'_n-x_n,\mu'_l-\mu_l)$ then the
differential $d{\bf z}$ evolves as,
\begin{equation}
d \dot{\bf z} = J d {\bf z},
\label{linflow}
\end{equation} 
where the Jacobian matrix $J$ is given by,
\begin{eqnarray}
J =
\left(
\begin{array}{ccccc}
{\partial f_1\over\partial z_1}-\epsilon & \partial f_1\over\partial z_2 & 
\cdots & \partial f_1\over\partial z_n & \partial f_1\over\partial
\mu'_l\\
\partial f_2\over\partial z_1 & \partial f_2\over\partial z_2 & 
\cdots & \partial f_2\over\partial z_n & \partial f_2\over\partial
\mu'_l\\
\vdots & \vdots & & \vdots & \vdots\\
\partial f_n\over\partial z_1 & \partial f_n\over\partial z_2 &
\cdots & \partial f_n\over\partial z_n & \partial f_n\over\partial
\mu'_l\\
{\partial h\over\partial z_1} & {\partial h\over\partial z_2} & 
\cdots & 
{\partial h\over\partial z_n} & {\partial h\over\partial \mu'_l}
\end{array}
\right),
\label{jacob}
\end{eqnarray}
where the function $h$ describes the evolution of the parameter $\mu_l'$
as in Eq.~(\ref{genadapt}) and the derivatives in the matrix $J$ are
evaluated at ${\bf z}=0$ which is a fixed point. The condition for the
convergence of our procedure is that the real part of the eigenvalues of
the matrix $J$ or {\it conditional Lyapunov exponents} (CLE's) are all
less than zero.

It can be seen from the above matrix equation that choices of the feedback
constant and the stiffness constant affect the values of conditional
Lyapunov exponents. Thus the method will work convergently only for
suitably chosen $\epsilon$ and $\delta$. When these are chosen such that
the largest of the CLE's become positive, the algorithm does not work due
to diverging trajectories.

In the next section we will illustrate the method using the examples of
Lorenz and R\"ossler flows and a set of equations in plasma physics.

\section{Illustrative Examples} 
\subsection{Lorenz system}

As a first example, we study the Lorenz system. We divide the discussion
in two parts. In the first, we present the results when only a single
parameter is estimated in a Lorenz system. Three different cases are
discussed in detail. In the later part, we extend our method for the case
when more parameters are to be estimated.

\subsubsection{Single parameter estimation}

The Lorenz system is given by,
\begin{eqnarray}
\dot x &= f_1(x,y,z) =& \sigma (y-x), \nonumber
\\
\dot y &= f_2(x,y,z) =& rx - y - xz, \nonumber 
\\ 
\dot z &= f_3(x,y,z) =& xy - bz, 
\label{lorenz1}
\end{eqnarray}
where $(x,y,z)$ form the state space and $(\sigma, r, b)$ form the three
dimensional parameter space. Now assume that the time series for $x$ is
given, and two of the three parameters are also known. We consider the
following cases. 

\noindent Case 1: When the unknown parameter appears in the
evolution of $x$:

Here assuming $\sigma$ to be the unknown parameter, we create a receiver
system as described in the Section I, given by,
\begin{eqnarray}
\dot x' &= g(x',y',z') =&\sigma'(y'-x')-\epsilon(x'-x(t)),\nonumber 
\\
\dot y' &= f_2(x',y',z') =& rx' - y' - x'z',\nonumber 
\\ 
\dot z' &= f_3(x',y',z') =& x'y'- bz' ,
\label{lorenz2}
\end{eqnarray}
where $(x',y',z')$ are the new state variables and $(\sigma',r,b)$ are
the parameters, $r$ and $b$ being the same as those in the transmitter
while $\sigma'$ is initially set to an arbitrary value. $\epsilon$ is the
the feedback constant. These constitute the receiver system. $x'$ is the
{\it drive} variable. 

The parameter $\sigma'$, which is initially set to an arbitrary value, is
made to evolve through an equation similar the equation
(Eq.~(\ref{adaptation})). Here we can use only the $\rm sign$ of the last
factor in Eq.~(\ref{adaptation}) since there is a single equation
involving parameter evolution.
\begin{eqnarray} 
\dot{\sigma'} = -\delta \left(x' - x(t)\right) {\rm sign}(y' - x').
\label{sprimedot}
\end{eqnarray}
This equation along with the receiver system (Eq.~(\ref{lorenz2})), can
achieve required synchronization as well as parameter estimation since, a
randomly chosen initial vector $(x',y',z')$ evolves to $(x,y,z)$ and
$\sigma'\rightarrow\sigma$ as time $t\rightarrow\infty$. 

Figure~1 displays the manner in which the synchronization takes place and
how the parameter $\sigma'$, initially set to an arbitrary value finally
evolves towards the precise ``unknown'' value $\sigma$. In Fig.~1(a), (b)
and (c)  we show the differences $x'-x , y'-y, z'-z$ as functions of time
and we observe that they eventually settle down to zero after an initial
transient. In Fig.~1(d) we plot $\sigma'-\sigma$ as a function of time
which also goes to zero simultaneously. 

The synchronization as shown in Fig.~1 occurs when conditional Lyapunov
exponents for the receiver system coupled to the parameter evolution are
all negative or at most zero. This restricts the suitable choices for
$\epsilon$ and $\delta$. The Jacobian matrix $J$, for the evolution of
the vector $(x'-x,y'-y,z'-z,\sigma'-\sigma)$ is given by
(Eq.~(\ref{linflow})),
\begin{eqnarray}
J =
\left(
\begin{array}{cccc}
-\sigma-\epsilon & \sigma & 0 & y'-x' \\
r-z' & -1 & -x' & 0 \\
y' & x' & -b & 0 \\
-\delta {\rm sign}(y'-x') & 0 & 0 & 0
\end{array}
\right).
\label{jacob1}
\end{eqnarray}

Figure~2 shows the curve along which the largest CLE becomes zero, in the
$(\epsilon,\delta)$ plane. In region I, all nontrivial CLE's are negative
and the method works convergently, while in region II, the largest CLE
becomes positive and no convergence takes place. Nevertheless note that
for any positive value of $\delta$, there can always be a suitably chosen
$\epsilon$ such that the convergence occurs. On the other hand, there is
a critical value of $\epsilon$ below which the method does not work.

\noindent Case 2 a : When the unknown parameter appears in the
evolution of $y$ variable :

Here, we consider the case of $r$ as the unknown parameter
(\ref{lorenz1}) and reconstruct the receiver as,
\begin{eqnarray}
\dot x' &= g(x',y',z') =&\sigma(y'-x')-\epsilon(x'-x(t)),\nonumber 
\\
\dot y' &= f_2(x',y',z') =& r'x' - y' - x'z',\nonumber 
\\ 
\dot z' &= f_3(x',y',z') =& x'y' - b z',
\label{lorenz3}
\end{eqnarray}
while the evolution of $r'$ takes the form, (Eqs.~(\ref{adaptation}) 
and (\ref{signcaln1})). Similar to Eq.~(\ref{sprimedot}) we use only the 
$\rm sign$ of the derivative involved.
\begin{eqnarray} 
\dot{r'} = -\delta \left(x'-x(t)\right) {\rm sign}(\sigma x'). 
\label{rprimedot}
\end{eqnarray}

When a time series for $x$ from (\ref{lorenz1}) is fed into the these
equations, setting $x',y',z'$ and $r'$ to arbitrary initial condition,
they finally evolve to the corresponding values of $x,y,z$ and $r$. The
associated Jacobian matrix (Eq.~(\ref{linflow}) is given by,
\begin{equation}
J =
\left(
\begin{array}{cccc}
-\sigma-\epsilon & \sigma  & 0 & 0 \\
r-z' & -1 & -x' & x' \\
y' & x' & -b & 0 \\
-\delta{\rm sign}(\sigma x') & 0 & 0 & 0
\end{array}
\right).
\label{jacob2}
\end{equation}
Figure~3 shows the curve along which the largest CLE becomes zero, in the
$(\epsilon,\delta)$ plane. In region I, all nontrivial CLE's are negative
and the method works convergently, while in region II, the largest CLE
is positive.

Let $\tau$ denote the time required for the convergence to the correct
value of the parameter within a given accuracy, defined as $A = (r'-r)/r$. 
In Fig.~4 we plot ($\tau$) as a function of the feedback constant
$\epsilon$, when the stiffness constant $\delta$ is held fixed. On the
other hand, $\tau$ may be plotted as a function of $\delta$ for a fixed
value of $\epsilon$.  This is plotted in Fig.~5. In both Fig.~4 and
Fig.~5, $r$ is assumed to be unknown and a time series for $x$ is assumed
to be given.  The chosen accuracy for convergence was $10^{-7}$.

In Fig.~6 we plot the time required for convergence of $r'$ to $r$ to
within a given accuracy as a function of logarithm of the accuracy, which
is 0 with respect to the initial value. The straight line shows that the
time required to achieve better accuracy grows exponentially. The slope of
the line in Fig.~6 corresponds to the Lyapunov exponent. It was compared
with the Lyapunov exponent computed using a numerical algorithm and a fair
agreement was observed.

\noindent Case 2 b : When the unknown parameter appears in the
evolution of $z$ variable :

The case where the parameter $b$ appearing in the evolution of $z$,
(Eq.~(\ref{lorenz1})) is unknown, while the given time series is for $x$
is a particularly interesting case. Since the variable $z$ does not
appear explicitly in the evolution equation for $x$,
the calculation of ${\rm sign}$ in
Eq.~(\ref{adaptation}) has to be done using Eq.~(\ref{signcaln2}). Thus
with the evolution for $b'$, the complete receiver system becomes,
\begin{eqnarray}
\dot x' &=& g(x',y',z') = \sigma(y'-x')-\epsilon(x'-x(t)),\nonumber 
\\
\dot y' &=& f_2(x',y',z') =  rx' - y' - x'z',\nonumber 
\\ 
\dot z' &=& f_3(x',y',z') =  x'y' - b' z', 
\label{lorenz4}
\\
\dot b' &=& -\delta \left(x'-x(t)\right) {\rm sign}(\sigma x' z'). 
\label{bprimedot}
\end{eqnarray}

An initial vector $(x',y',z',b')$ in the above system goes to 
$(x,y,z,b)$ and thus makes the estimation of the value of $b$
possible. Here the matrix $J$ takes the form, (Eq.~(\ref{linflow}))
\begin{equation}
J =
\left(
\begin{array}{cccc}
-\sigma-\epsilon & \sigma  & 0 & 0 \\
r-z' & -1 & -x' & 0 \\
y' & x' & -b & -z' \\
-\delta{\rm sign}(\sigma x' z') & 0 & 0 & 0
\end{array}
\right).
\label{jacob3}
\end{equation}
Figure~7 shows the curve along which the largest CLE becomes zero in the
$\epsilon-\delta$ plane. In region(I), all CLE's are
negative and the condition of convergence is satisfied. 

Finally we note that in all the three cases discussed above, since the
time series for $x$ in Eq. (\ref{lorenz1})  is assumed to be known, $x'$
acts as a drive variable. A similar procedure is possible when a time
series for $y$ in Eq.~(\ref{lorenz1}) is given as an input. Here $y'$ can
be chosen as a drive variable which drives the evolution of the remaining
variables as well as the unknown parameter. Thus it is possible to know
an unknown value of any of the parameters of the Lorenz system from a
single time series for $x$ or $y$.

\subsubsection{Extension to many parameters' estimation}

Here we will consider the estimation of two or three parameters for the
Lorenz system (\ref{lorenz1}). 

We have applied our method for estimation of two parameters of the Lorenz
system (\ref{lorenz1}), taking $x$ or $y$ as drive variables. A
typical receiver system, taking $x$ as the drive and $(\sigma,r)$ as the
unknown parameters, is constructed as, 
\begin{eqnarray}
\dot x' &=&\sigma'(y'-x')-\epsilon(x'-x(t)),\nonumber 
\\
\dot y' &=& rx' - y' - x'z',\nonumber 
\\ 
\dot z' &=& x'y'- bz' ,\nonumber
\\
\dot{\sigma'} &=& -\delta \left(x' - x(t)\right)(y'-x'),\nonumber
\\
\dot{r'} &=& -\delta \left(x'-x(t)\right) (\sigma x'). 
\label{twoparest}
\end{eqnarray}
Note that the same stiffness constant is used in controlling both the
unknown parameters.

We have found that with similar receiver structure to that in
Eq.~(\ref{twoparest}), it is possible to estimate any two of the three
parameters $\sigma, r$ and $b$, when a time series for either of $x$ or
$y$ is given. In Fig.~8 (a) we plot the difference $(\sigma'-\sigma)$
while Fig.~8 (b) shows $(r'-r)$ as functions of time, when the drive is
$x$ and two parameters $\sigma$ and $r$ are assumed unknown to the
receiver. We see that the differences converge to zero, indicating
that it is possible to estimate two parameters simultaneously. 

Finally we mention that, if the time series for $y$ is given, estimation
of all the three parameters is possible though in this case, the
convergence is very slow. The method fails to estimate all the three
parameters $\sigma, r$ and $b$, when time series for $x$ is given.

We thus note that the detailed information about all the parameters of
Lorenz system is contained in a time series for either $x$ or $y$
variables and can be extracted as above.

It should be however mentioned that when a time series for $z$ is given
from a Lorenz system, the eigenvalues of the associated matrix $J$
(Eq.~(\ref{jacob})) do not satisfy the condition of convergence for any
choice of $\epsilon$ and $\delta$. Thus the method fails when a time
series for $z$ is known. 

\subsection{R\"ossler system}

We next consider the R\"ossler system of equations given by,
\begin{eqnarray}
\dot x &=& -y-z,\nonumber
\\ 
\dot y &=& x + ay,\nonumber
\\ 
\dot z &=& b + z(x-c).
\label{rossler1}
\end{eqnarray}
which contains the three parameters $(a,b,c)$.

We have applied our procedure to estimate any of these parameters, when
unknown, assuming the knowledge of a time series for the variable $y$ in
the R\"ossler system. The corresponding variable $y'$, which acts as a
drive variable for $(x',y',z')$ and the evolution of the unknown
parameter, then evolves through,
\begin{equation}
\dot y' = x' + ay' - \epsilon (y'-y(t)),
\end{equation}
while the unknown parameter evolves adaptively. 

Thus with the given time series $y(t)$, fed into the evolution of the
drive variable $y'$, we find that the convergence condition can be
satisfied by a suitable choice of feedback constant and the stiffness
constant. 

In Fig.~9 we show the convergence  of ($x'-x, y'-y, z'-z, a'-a$) 
to (0, 0, 0, 0) when the parameter $a$ is
unknown. Thus our algorithm of parameter estimation works for $y'$ as a
drive variable and any of the three parameters can be estimated. We have
however found that the convergence is not possible for $x$ or $z$ as the
drive variables. 

Finally, we have also applied our method to estimate two or three
parameters of the Rossler system with $y$ as a drive variable. It was seen
that no choice of the feedback constant and the stiffness constant lead to
convergences required for estimation.

\subsection{An example from plasma physics} 

As our final example, we present a set of nonlinear equations appearing
in plasma physics. This is the so called resonant three-wave coupling
equations when high frequency wave is unstable and the remaining two are
damped~\cite{Wer}. These equations are,
\begin{eqnarray}
\dot a_1 &=& a_1 + {a_2}^2 {\rm cos}\phi, \nonumber \\
\dot a_2 &=& -a_2(\gamma + a_1 {\rm cos}\phi), \nonumber \\
\dot \phi &=& -\delta + a^{-1}_1(2{a_1}^2 - {a_2}^2){\rm sin}\phi,
\label{plasma1}
\end{eqnarray}
where $\gamma$ and $\delta$ are the system parameters.  

We find that with time series given for either $a_1$ or $a_2$, it is
possible to know an unknown parameter $\gamma$ or $\delta$ using
synchronization and adaptive control. The method fails when a time series
for $\phi$ is known.

Figure~10 displays the evolution of the differences between the
transmitter and receiver variables as well as the evolution of
$\delta'-\delta$ as functions of time, when the time series for $a_1$ is
known. As expected, the differences go to zero asymptotically.

\section{Effect of noise}

In this section we will study the effect of noise present in the
transmitter system. We will take the example of the Lorenz system
(Eq.~(\ref{lorenz1}) for this purpose, where $\sigma$ is assumed to be
the unknown parameter and $x$ acts as a drive variable. 

Assuming that there is a small additive noise present in the time series
given for $x$, we feed the noisy time series into the receiver system
(Eq.~(\ref{lorenz2}) and carry out the parameter estimation as described. 

We find that for weak noise, the asymptotically estimated value of the
parameter fluctuates around the correct value with a small amplitude.  
Thus the estimation is possible using our method. The error in the
estimation can be reduced by a suitable averaging over the time evolution
of $\sigma'$ in the asymptotic limit. For increasing strengths of noise,
the fluctuations in the estimated value grow larger and precise estimation
becomes difficult. Figure~11 shows the convergence of $\sigma'$ to
$\sigma$ when additive noise is present in the evolution of $x$, the drive
variable of the Lorenz system, for which the time series is given.

We define the accuracy ($A$) in the estimation of $\sigma$ as
$(\sigma'-\sigma)/\sigma$ while $w$ denotes the strength of noise with
uniform distribution ranging from $-w$ to $w$. In Fig.12 we plot the
asymptotic value of A, the accuracy of the estimation of $\sigma$, against
the strength $w$ of noise in $x$. It can be seen from the curve that the
accuracy grows linearly as the noise increases to a value of
$w = 2$ which corresponds to about 12\% of the range of $x$ values.  The plot
thus shows that our method is quite robust for weak noise in
$x$, while it can fail as the noise strength increases to a larger value.

\section{Conclusions}

To summarize, we have shown that a combination of synchronization based
on linear feedback given into only a single receiver variable with an
{\it adaptive} evolution for parameters unknown to the receiver, enables
the estimation of the unknown parameters.  The feedback comes from a
scalar time series. We have also shown that our procedure corresponds
to dynamic minimization of the synchronization error.

We have presented examples of Lorenz and R\"ossler systems taking
different candidate parameters to be unknown to the receiver as well as
that of a plasma system obeying resonant three-wave coupling equation. In
the Lorenz system (Eq.~(\ref{lorenz1})), any of the three parameters can
be estimated when a time series is given for either of $x$ and $y$, but
the method fails when the known time series is for the variable $z$. 
Extensions to estimation of more than one parameters of the Lorenz system
are also presented as a representative case. Estimation of two parameters
is possible for both $x$ or $y$ as drive variables while estimation of all
the three parameters is possible only when time series for $y$ is given.  

In the case of R\"ossler system (Eq.~(\ref{rossler1}))  the method works
only when the time series is given for the variable $y$ where it is
possible to estimate any of the three parameters.  We find that in case of
the plasma system, the parameters can be estimated with the feedback in
the evolution for either $a_1$ or $a_2$.

We have thus numerically demonstrated that the explicit detailed
information about the parameters of a nonlinear chaotic system is
contained in the time series data of a variable and can be extracted under
suitable conditions. This information includes the particular values of
the parameters of the system which can be estimated even if they
appear in the evolution of variables other than the one for
which the time series is given. 

We have also checked the robustness of the method against the noise and
it shows reasonable robustness against small noise though the error of
estimation becomes larger as the noise strength is increased. 

The possibility of improving the efficiency of the method needs to be
explored. This can be done, for example, by optimizing the choices of
newly introduced parameters $\epsilon$ and $\delta$ or by trying to
estimate initial values of variables of the transmitter system,
corresponding to response variables and thereby starting from a ``better''
initial point. Work in these directions is under progress.

One of the authors(AM) will like to thank UGC, India and the other (REA)
will like to thank DST, India for financial support.


\newpage
\centerline{\huge Figure Captions}

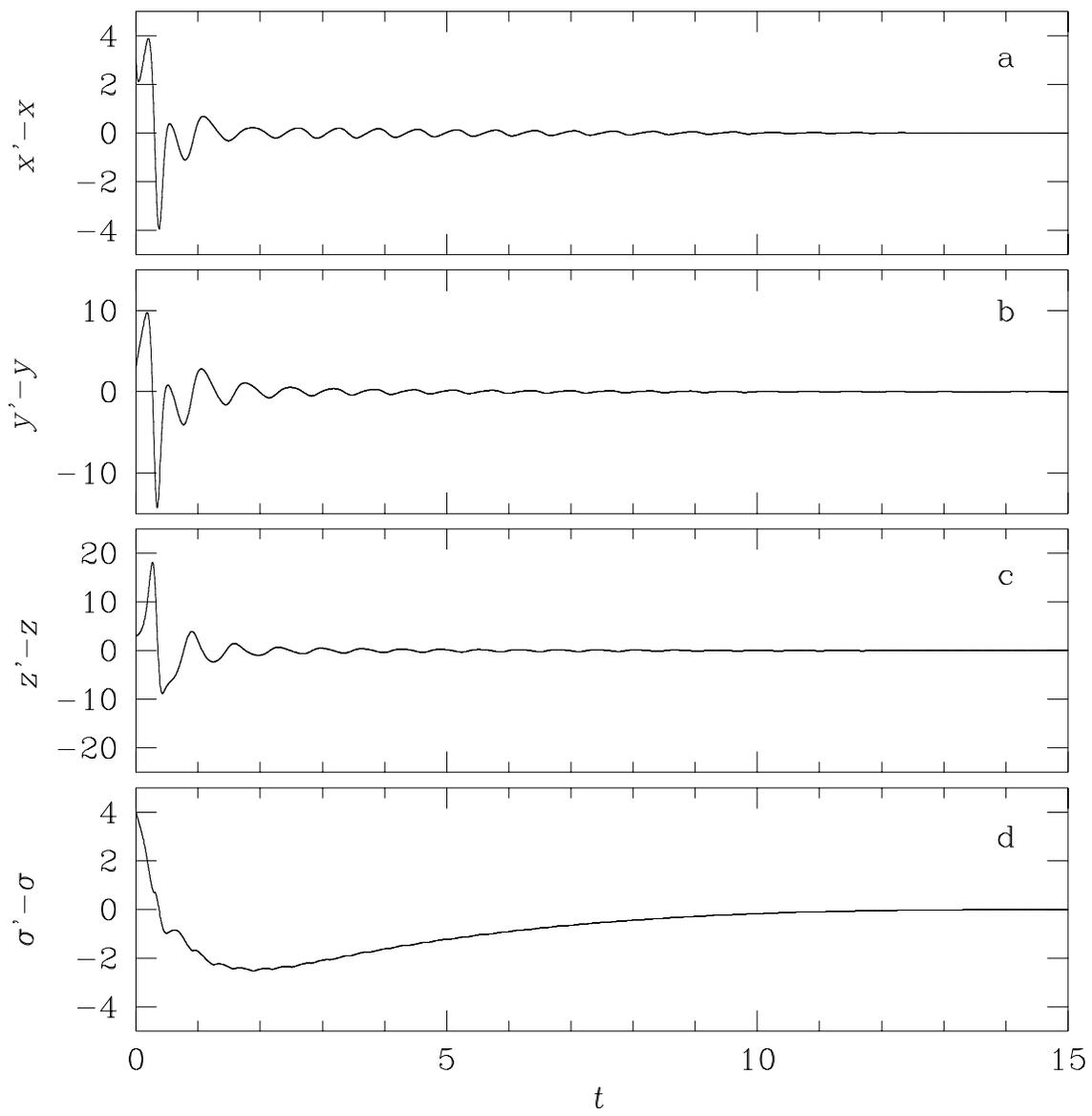
FIG.1. Figures (a), (b), (c) and (d) show the differences $(x'-x, y'-y,
z'-z, \sigma'-\sigma)$ respectively as functions of time, for the Lorenz
system ~(Eqs.(\ref{lorenz1}), (\ref{lorenz2}) and (\ref{sprimedot})).  The
unknown parameter is $\sigma$ and the drive variable is $x$.  The figures
show that the differences tend to zero asymptotically. $\sigma'$ which is
set to an arbitrary initial value finally evolves to $\sigma$ facilitating
the parameter estimation to any desired accuracy in the asymptotic limit. 

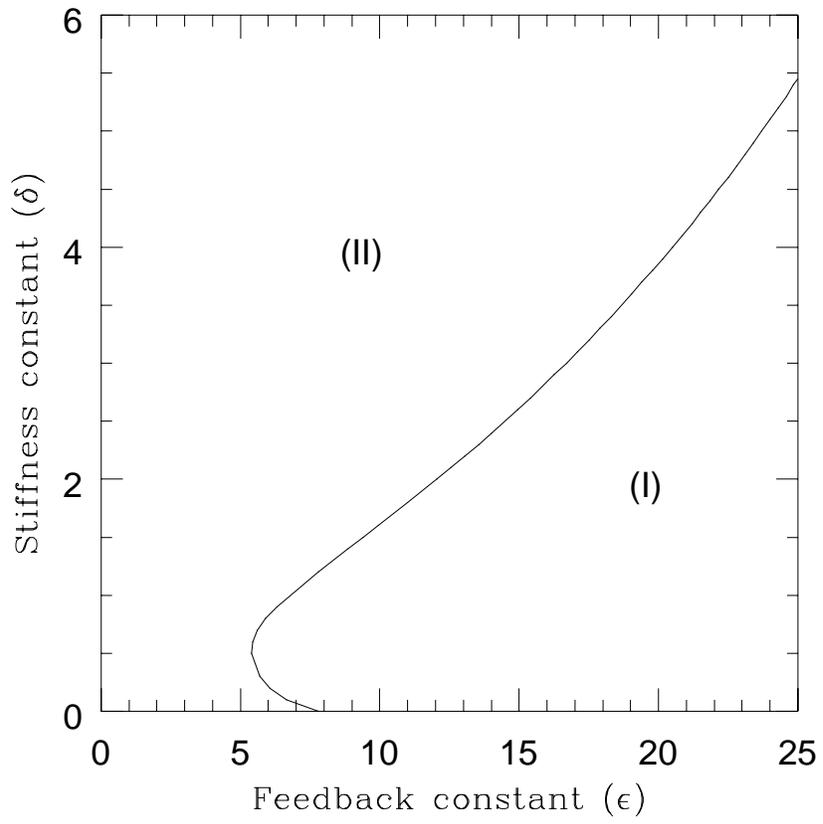
FIG.2. The curve along which the largest conditional Lyapunov exponent
(computed using Eq.~(\ref{jacob1})) becomes zero in the
$(\epsilon, \delta)$-plane for the Lorenz system with $\sigma$ as the
unknown parameter (Eqs.~(\ref{lorenz2}) and (\ref{sprimedot}) is plotted. 
In region (I), the CLE's are all negative and parameter estimation works
convergently. Region (II) corresponds to a positive largest CLE, where the
method does not work. Note that there is a critical $\epsilon$ below which
the method does not work. Nevertheless for any $\delta$, an $\epsilon$ can
be chosen so that the method works. 

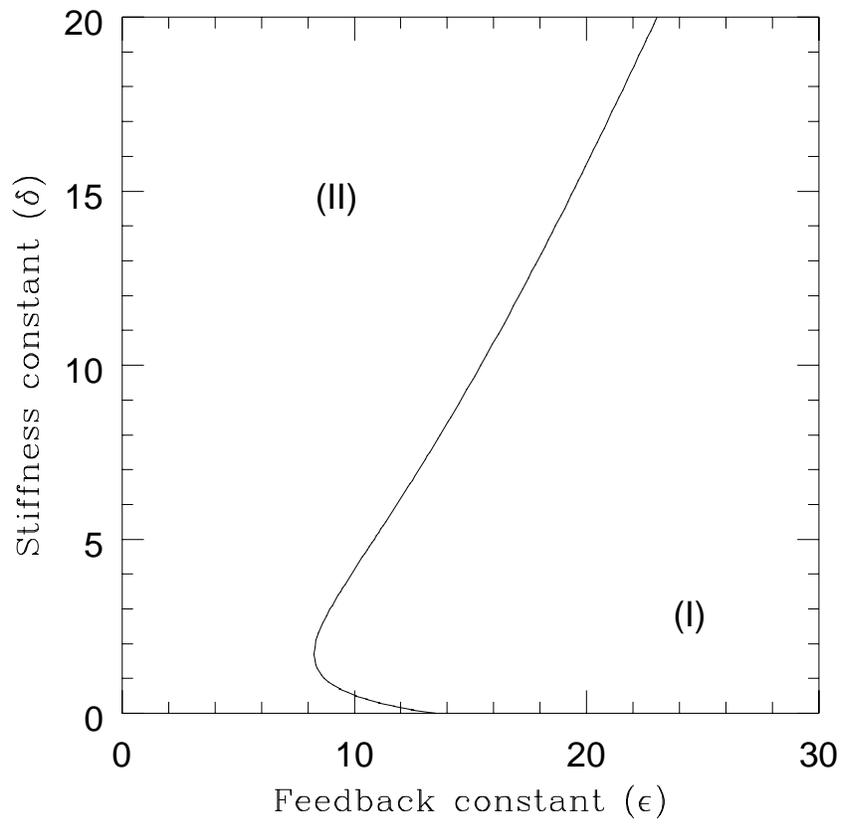
FIG.3. The figure shows the curve along which the largest conditional
Lyapunov exponent for Lorenz system with the parameter $r$ as unknown and
$x$ as drive variable (Eqs.~(\ref{lorenz2}) and (\ref{rprimedot}) becomes
zero in the $(\epsilon, \delta)$-plane. In region (I) all the CLE's are
negative and the parameter estimation can be achieved. In region (II) the
the largest Lyapunov exponent is positive.

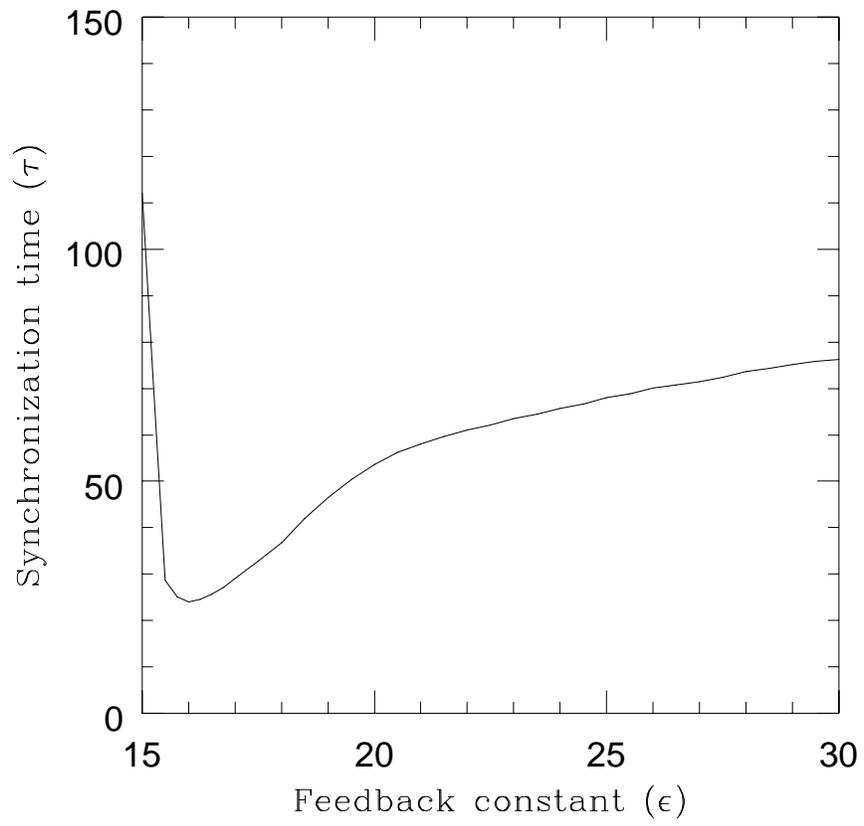
FIG.4. The plot shows the time ($\tau$) required for convergence of $r'$
to $r$ to a given accuracy with a fixed value of the stiffness constant
($\delta$), as a function of the feedback constant, $\epsilon$, for Lorenz
system.  The drive is $x$ while the unknown parameter is $r$
(Eq.~(\ref{lorenz2})). It can be seen that the synchronization time tends
to infinity when the largest CLE becomes zero. 

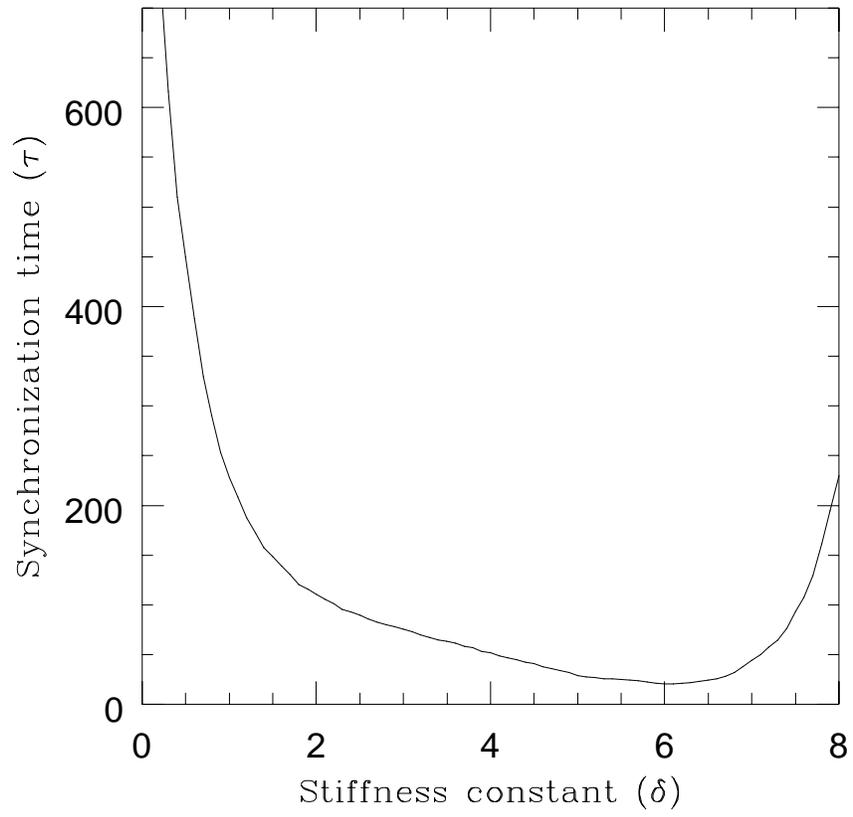
FIG.5. The plot shows the time ($\tau$) required for convergence of $r'$
to $r$ 2 a given accuracy, with a fixed value of the feedback constant
($\epsilon$), as a function of the stiffness constant, $\delta$, for the
Lorenz system.  The drive is $x$ while the unknown parameter is $r$
(Eq.~(\ref{lorenz2})). It can be seen that the synchronization time tends
to infinity as $\delta$ approaches a value so as to make the largest CLE
zero. 

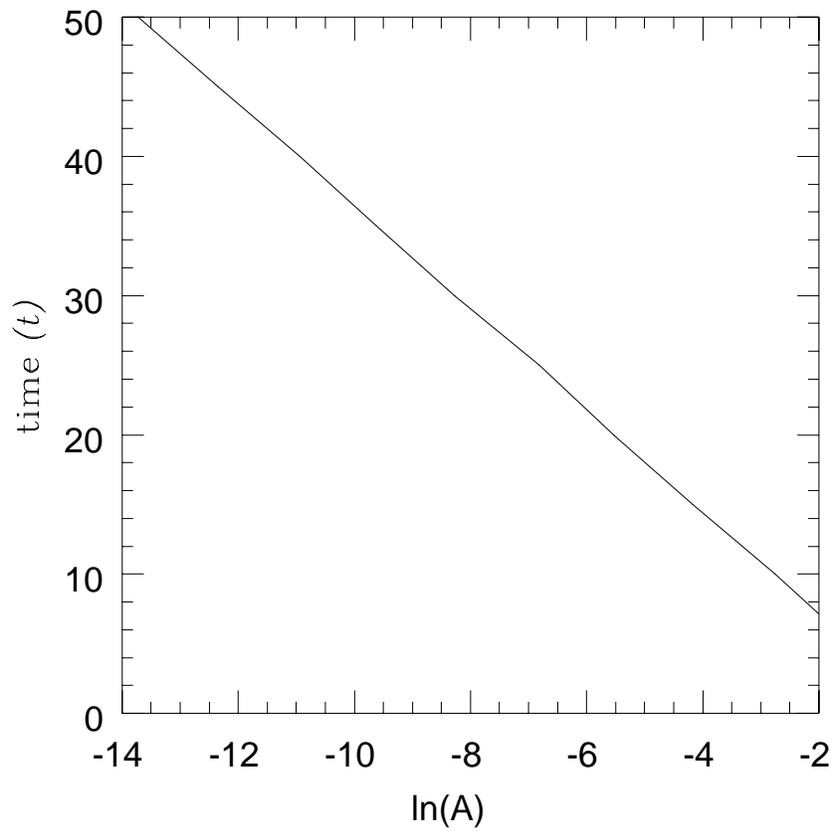
FIG.6. The graph shows the time, t, required to achieve the parameter
estimation to within a given accuracy as a function of the accuracy, A,
(logarithmic scale) normalized with respect to the initial deviation of
the parameter from the correct value for Lorenz system
(Eq.~(\ref{lorenz2})). The time series for $x$ is assumed to be known
while the value of $r$ is unknown. The straight line shows that the time
required for a better accuracy grows exponentially.

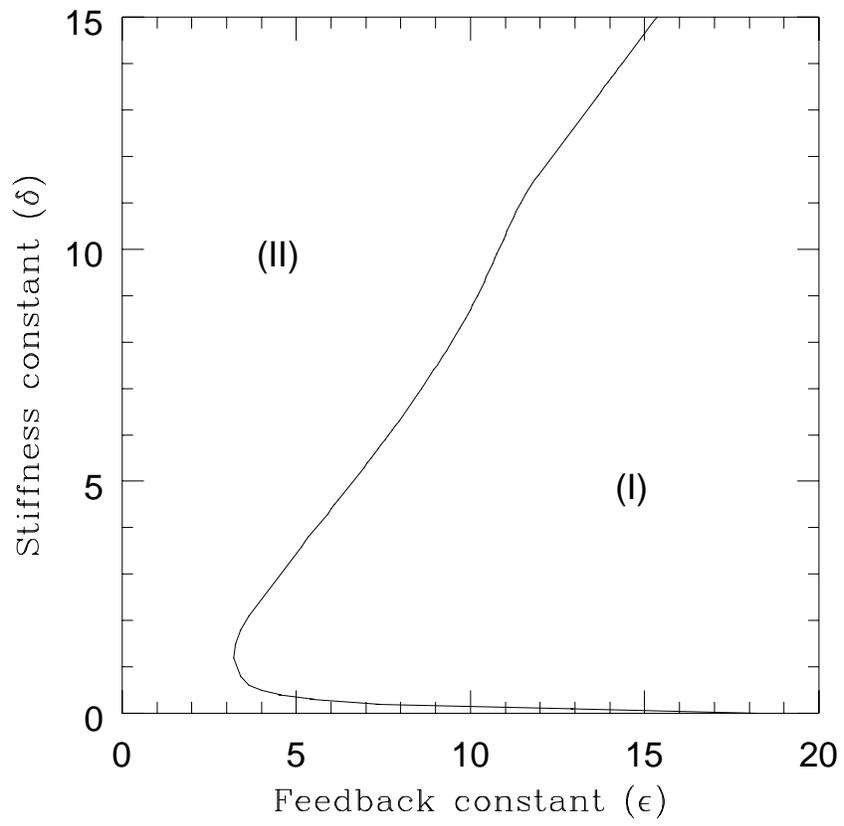
FIG.7. The curve along which the largest conditional Lyapunov exponent
(computed using Eq.~(\ref{jacob1})) becomes zero in the $(\epsilon,
\delta)$-plane for the Lorenz system with $b$ as the unknown parameter and
$x$ as drive (Eqs.~(\ref{lorenz4}) and (\ref{bprimedot}) is plotted. In
region (I), the CLE's are all negative and parameter estimation works
convergently. Region (II) corresponds to a positive largest CLE, where the
method does not work. Similar to other cases, there is a critical
$\epsilon$ below which the method does not work.

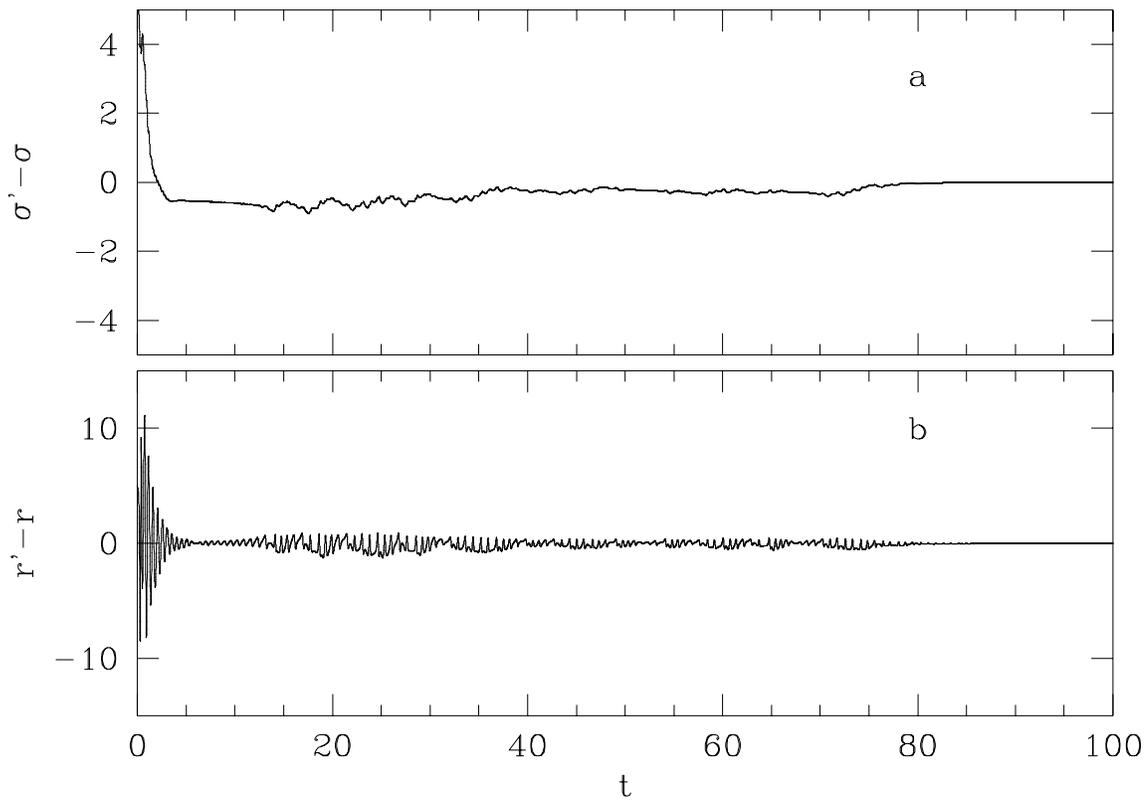
FIG.8. Plots (a) and (b) show the differences $(\sigma'-\sigma)$ and
$(r'-r)$ respectively as
functions of time in the Lorenz system~(Eq.(\ref{lorenz1}).
The unknown parameters are $\sigma$ and $r$ and the drive variable is $x$.
The plots show that the differences go to zero, and hence indicate that a
simultaneous estimation of more than one unknown parameters is possible.

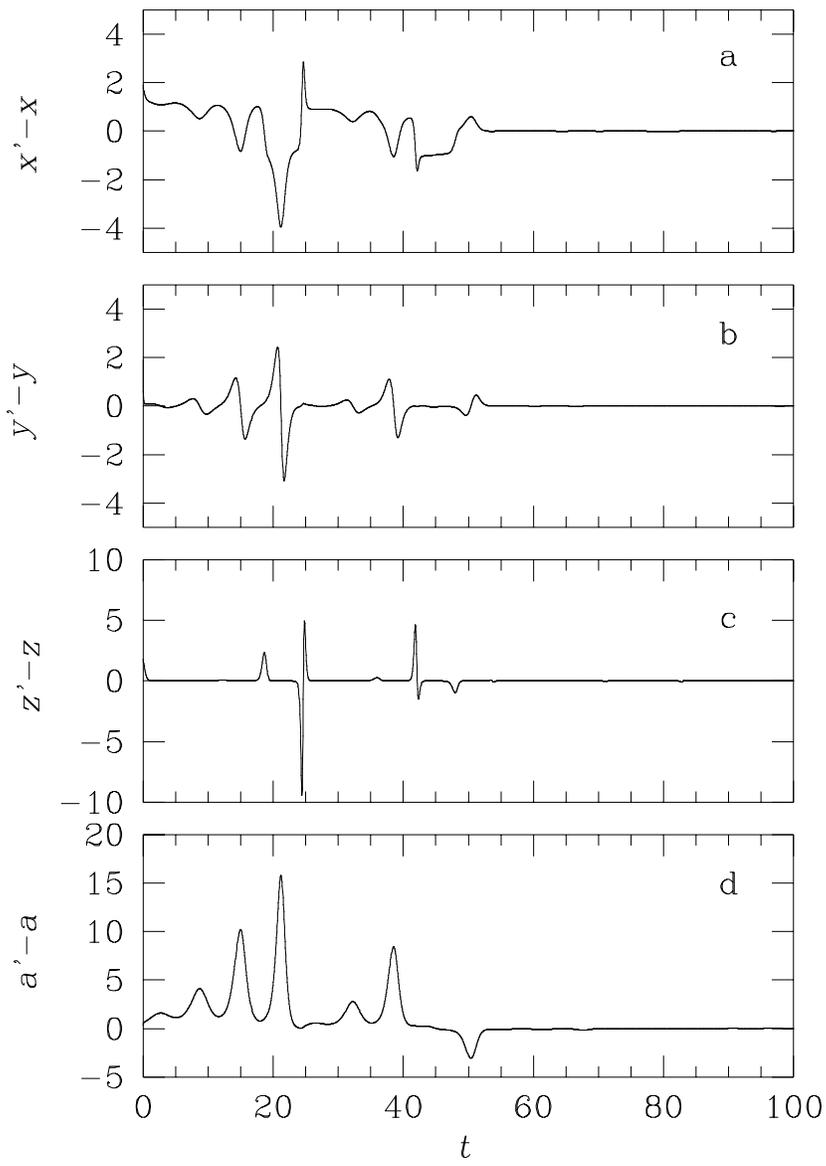
FIG.9. Plots (a), (b), (c) and (d) show the differences $(x'-x, y'-y,
z'-z, a'-a)$ as functions of time respectively, in the R\"ossler system
~(Eq.(\ref{rossler1})). The unknown parameter is $a$ and the drive
variable is $y$. The figures show that the differences tend to zero
asymptotically. $a'$ which is set to an arbitrary initial value finally
evolves to $a$ facilitating the parameter estimation. 

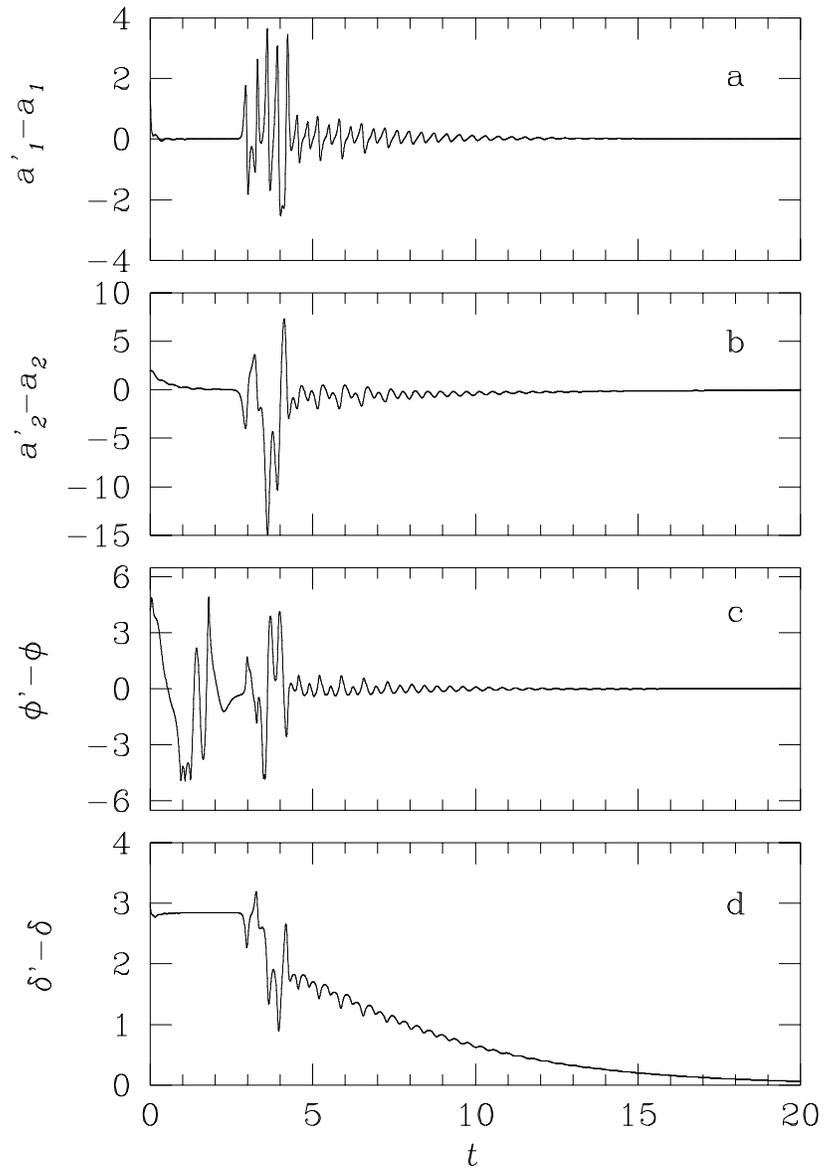
FIG.10. Plots (a), (b), (c) and (d) show the differences
$(a'_1-a_1, a'_2-a_2, \phi'-\phi, \delta'-\delta)$ as functions of time
respectively, in the plasma system ~(Eq.(\ref{plasma1})). The unknown
parameter is $\delta$ and the drive variable is $a_1$. The figures show
that the differences tend to zero asymptotically. $\delta'$ which is set
to an arbitrary initial value finally evolves to $\delta$ facilitating the
parameter estimation. 

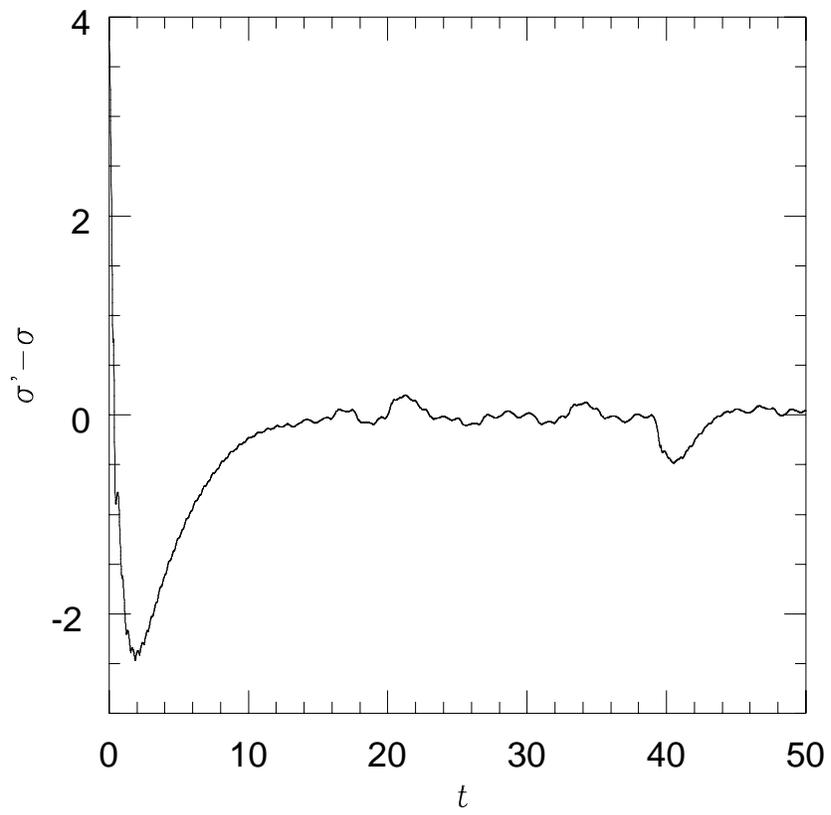
FIG.11. The graph shows the evolution of $\sigma'-\sigma$ as a function of
time, in the presence of an additive noise $(w = 0.1)$ in the given time
series for $x$ for Lorenz system (\ref{lorenz1}). The value of $\sigma$ is
assumed unknown. The plot shows that the difference $\sigma'-\sigma$ fluctuates
around zero with a small amplitude after an initial transient and a
reasonably good estimation is possible using a suitable averaging over
these fluctuations.

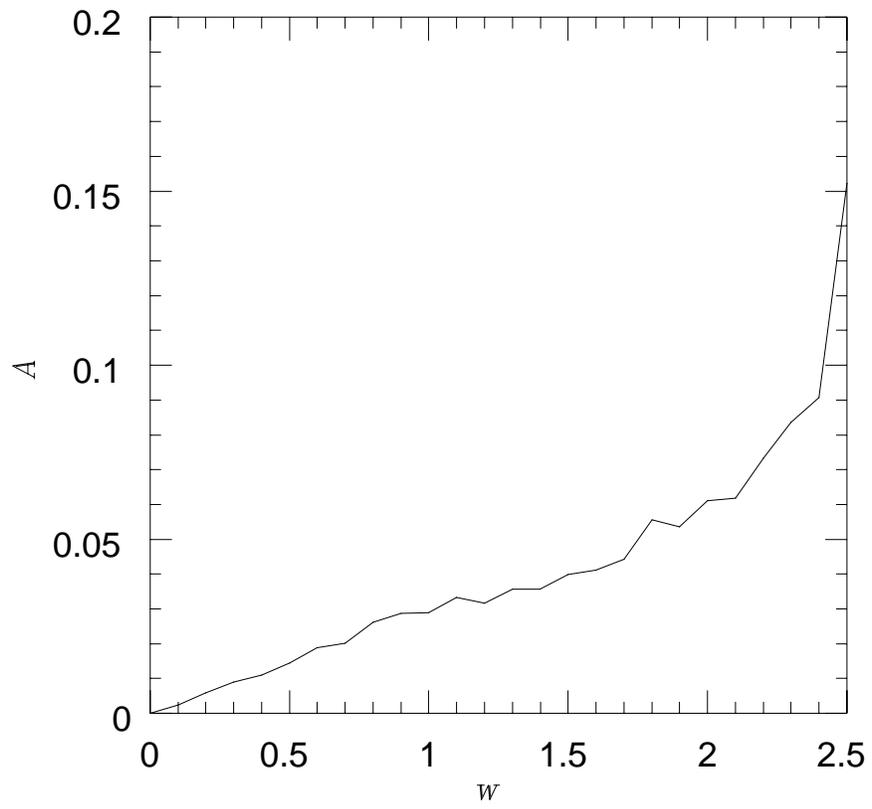
FIG.12. The plot of asymptotic accuracy of parameter estimation
($A=(\sigma'-\sigma)/\sigma$), as a function of strength of the noise,
$w$, in the given time series of $x$ in Lorenz system
(Eq.~(\ref{lorenz1})). The noise with strength $w$ takes uniformly
distributed values from $-w$ to $+w$. The drive is $x$ and the unknown
parameter is $\sigma$. It is seen that the estimation of $\sigma$ is
stable for a range of noise strength growing from zero to about 2 which
corresponds to about 12 \% of the range of $x$ values. 
\end{document}